\documentclass[prc, aps, twocolumn,showkeys, showpacs,amsmath,amssymb,superscriptaddress]{revtex4-1}
\usepackage{graphicx}
\usepackage{dcolumn}
\usepackage{bm}
\usepackage{epsfig}
\usepackage{subfigure}
\usepackage{color}

\begin{document} 
\title{On the relativistic Lattice Boltzmann method for quark-gluon
  plasma simulations}

\author{D. Hupp} \email{huppd@student.ethz.ch} \affiliation{ ETH
  Z\"urich, Computational Physics for Engineering Materials, Institute
  for Building Materials, Schafmattstrasse 6, HIF, CH-8093 Z\"urich
  (Switzerland)}

\author{M. Mendoza} \email{mmendoza@ethz.ch} \affiliation{ ETH
  Z\"urich, Computational Physics for Engineering Materials, Institute
  for Building Materials, Schafmattstrasse 6, HIF, CH-8093 Z\"urich
  (Switzerland)}

\author{I. Bouras} \email{bouras@th.physik.uni-frankfurt.de}
\affiliation{ Institut f\"ur Theoretische Physik, Johann Wolfgang
  Goethe-Universit\"at, Max-von-Laue-Strasse 1, D-60438 Frankfurt am
  Main, (Germany)}

\author{S. Succi} \email{succi@iac.cnr.it} \affiliation{Istituto per
  le Applicazioni del Calcolo C.N.R., Via dei Taurini, 19 00185, Rome
  (Italy),\\and Freiburg Institute for Advanced Studies,
  Albertstrasse, 19, D-79104, Freiburg, (Germany)}

\author{H. J. Herrmann}\email{hjherrmann@ethz.ch} \affiliation{ ETH
  Z\"urich, Computational Physics for Engineering Materials, Institute
  for Building Materials, Schafmattstrasse 6, HIF, CH-8093 Z\"urich
  (Switzerland)} \affiliation{Departamento de F\'isica, Universidade
  Federal do Cear\'a, Campus do Pici, 60451-970 Fortaleza, Cear\'a,
  (Brazil)}

\date{\today}
\begin{abstract}
  In this paper, we investigate the recently developed lattice
  Boltzmann model for relativistic hydrodynamics.  To this purpose, we
  perform simulations of shock waves in quark-gluon plasma in the low
  and high viscosities regime, using three different computational
  models, the relativistic lattice Boltzmann (RLB), the Boltzmann
  Approach Multi-Parton Scattering (BAMPS), and the viscous sharp and
  smooth transport algorithm (vSHASTA).  From the results, we conclude
  that the RLB model departs from BAMPS in the case of high speeds and
  high temperature(viscosities), the departure being due to the fact
  that the RLB is based on a quadratic approximation of the
  Maxwell-J\"uttner distribution, which is only valid for sufficiently
  low temperature and velocity. Furthermore, we have investigated the
  influence of the lattice speed on the results, and shown that
  inclusion of quadratic terms in the equilibrium distribution
  improves the stability of the method within its domain of
  applicability.  Finally, we assess the viability of the RLB model in
  the various parameter regimes relevant to ultra-relativistic fluid
  dynamics.
\end{abstract}

\pacs{47.11.-j, 12.38.Mh, 47.75.+f}

\keywords{Lattice Boltzmann, quark-gluon plasma, relativistic fluid
  dynamics}

\maketitle

\section{Introduction}

Recent experiments on heavy-ion (Au-Au) collisions with
ultra-relativistic energies at the Relativistic Heavy Ion
Collider(RHIC) at Brookhaven National Laboratory, have revealed a new
state of matter, the quark-gluon plasma, whereby hadronic matter
undergoes a deconfining transition which liberates quarks in the form
of a gas of quasi-free particles \cite{qgp}.  The quark-gluon plasma,
which is credited for dominating the primordial state of the Universe
in its earliest $1-10$ microseconds, shows very interesting
properties, such as near-perfect fluid-like behavior, characterized by
ultralow dynamic shear viscosity and associated onset of shock wave
propagation \cite{qcshockwave, data}.  The study of such shock waves
plays a major role in the characterization of the quark-gluon plasma,
since they carry information both on its equilibrium (equation of
state) and non-equilibrium (transport coefficients) properties.

In the last years, several numerical tools have been used for the
computational investigation of quark-gluon plasmas, such as, for
instance: the Boltzmann approach of multiparton scattering (BAMPS)
\cite{BAMPS}, which solves the full Boltzmann equation,
$p^{\mu}\partial_{\mu}f(x,p) = C(x,p)$, with $p^\mu$ the microscopic
4-momentum vector, $f(x,p)$ the single particle distribution function,
and $C(x,p)$ the binary collision term; and the viscous sharp and
smooth transport algorithm (vSHASTA)\cite{vSHASTA}, which is based on
a hydrodynamic description.

\begin{figure}[]
  \centering
  \includegraphics[width=0.2\textwidth]{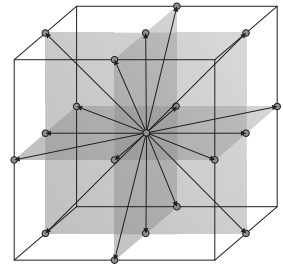}
  \caption{The D3Q19 (19 speeds in 3 dimensions) cell configuration. 
    Every arrow corresponds to the
    scaled velocity vector $\delta t \vec{c}_i$, and the points
    correspond to the discretized space coordinates $\vec{x}$.}
  \label{fig:D3Q19}
\end{figure}

Recently, the relativistic lattice Boltzmann (RLB) model \cite{RLB1}
was introduced. This new method is a relativistic extension of the
classical lattice Boltzmann (LB) equation\cite{Succi01}, whereby a
minimal form of the Boltzmann equation is discretized on a lattice and
the collision term is applied in a single relaxation time, the
so-called Bhatnagar-Gross-Krook approximation, Ref. \cite{BGK}). In
the above, "minimal form", implies that the particle velocities are
constrained to take only a very limited set of discrete values,
typically of order $10$ and $20$ in two and three dimensions,
respectively (see Fig.\ref{fig:D3Q19}), which represents an enormous
simplification as compared to the true Boltzmann equation.  The key is
that the proper selection of this handful of discrete speeds is
sufficient to compute {\it exactly} the low-order kinetic moments
which characterize the hydrodynamic regime.  Crucial to the success of
this procedure is that the system be only weakly out of equilibrium,
so that the particle distribution function remains close to a local
Maxwellian, whose low-order kinetic moments can be computed exactly
with just a few suitably chosen discrete velocities (the nodes of
Gaussian integration).

The LB method shows many advantages, as for example, the relatively
easy implementation of the simulations of fluids in complex
geometries, excellent suitability to parallel implementation, and high
flexibility towards the inclusion of additional physics, besides sheer
hydrodynamics \cite{complex1,complex2,complex3,complex4,complex5}.  It
can thus be expected that the RLB model would carry many of these
assets over to the relativistic context. To date, it has been shown to
compute shock-wave formation and propagation in quark-gluon plasmas at
a fraction of the cost of relativistic hydrodynamic codes
\cite{RLB1,RLB2}.

In order to characterize and improve the RLB model, further studies on
its capabilities, limitations and computational performance are in
order, which is precisely the focus of this paper. To this purpose, it
is worth reminding that the actual RLB model deals with weakly
relativistic fluids, characterized by $1<\gamma<2$ and $\zeta \equiv
\frac{mc^2}{k_bT}>1$, where $\gamma = (1-u^2/c^2)^{-1/2}$ is the
Lorentz factor associated with a fluid with speed $u$ and $T$ is the
fluid temperature.  However, since $\gamma=2$ corresponds to
$\beta=u/c \sim 0.85$, and $\zeta=1$ associates with ultra-high
temperatures, typically above $10^{13}$ Kelvin degrees, the weakly
relativistic regime embraces nonetheless a substantial number of
interesting applications, including the quark-gluon plasma generated
by recent experiments on heavy-ions and hadron jets \cite{QG-1, QG-2,
  QG-3, QG-4, QG-5, QG-6, QG-7}, as well as astrophysical flows, such
as interstellar gas and supernova remnants \cite{supernova2,
  supernova3}.

An extension of the original RLB model, capable of dealing with
non-Minkowskian geometries and ultra-relativistic fluids, has been
recently developed \cite{rlbPRC}.  However, since this method is
comparatively more elaborate than the original RLB, in the sequel we
shall confine our attention to the latter.

This work is organized as follows: first, in section~\ref{sec:imp} we
give a short introduction to the RLB model, and show some simulations
to validate our implementation. Subsequently, in section
~\ref{sec:improvement}, we extend the equilibrium distribution
function for the particle density, to obtain the correct density
profile in the moderately relativistic regime $\gamma \sim 2$.
Finally, in section ~\ref{sec:infcl}, we analyze the effect of the
lattice speed on the results, provide comparisons in the moderately
relativistic and high viscosity regimes with the previous mentioned
methods, BAMPS and vSHASTA, and assess the viability of this scheme
for ultra-relativistic fluid dynamics.

\section{Relativistic lattice Boltzmann model}
\label{sec:imp}
In this section, we describe the relativistic lattice Boltzmann, which
is the basis of this work, develop some improvements and test the new
implementation against known results for the Riemann problem.  The
Riemann problem has piecewise constant initial conditions and a single
discontinuity, and is commonly used to test numerical methods that
model systems with conservation laws.

\subsection{Conservation laws for relativistic fluid dynamics}

The conservation laws for relativistic fluid dynamics, which our RLB
model is based upon, are: the conservation of particle density (in a
relativistic regime mass is not invariant), and the energy momentum
conservation, which has to be treated separately in contrast to the
classical LB.

The conservation of the particle number density $n$ is given by:
\[ \partial_t\left(n\gamma\right) + \partial_a\left( n\gamma u_a
\right) = 0 \quad ,\] where $\vec{u}$ denotes the macroscopic fluid
velocity. Here, Latin indices denote the three dimensional space
coordinates.  Note that in the following the speed of light $c$ and
the Boltzmann constant $k_b$ are taken to be unity for notational
simplicity. As a result, $\beta=u$ and $\zeta=m/T$.  Also we use the
Einstein summation, i.e. repeated indices are summed upon.  The
hydrodynamic equations for the energy momentum conservation read as
follows:
\begin{equation}
  \begin{aligned}
    \partial_t \left(\left(\epsilon + P \right) \gamma^2 - P \right)
    &+ \partial_a\left(\left(\epsilon
	+ P \right)\gamma^2 u_a\right) \\ &+ \partial_t \pi^{00} + \partial_a \pi^{ab} = 0 \quad ,\\
    \partial_t \left(\left(\epsilon + P\right)\gamma^2 u_{ab}\right)
    &+
    \partial_b P + \partial_a\left(\left
	(\epsilon + P\right)\gamma^2 u_a u_b\right) \\ &+\partial_t \pi^{0b} + \partial_b \pi^{ab} = 0 \quad ,\\
  \end{aligned}
\end{equation}
where $\epsilon$ is the energy density, $P$ is the hydrostatic
pressure, and $\pi^{ab}$ are the components of the dissipation
stress-energy tensor. Here the $0$ index denotes the time component.

\subsection{Implementation of the RLB scheme}

We implement the relativistic lattice Boltzmann model along the lines
proposed in Refs.\cite{RLB1,RLB2}. This model is based on two density
distribution functions, $f_i$ and $g_i$, which are associated to a
D3Q19 lattice \cite{Succi01} (see Fig.~\ref{fig:D3Q19}), with $19$
discrete velocities $\vec{c}_i$ in $d=3$ dimensions.

These velocities take just the values $0, \pm c_l$, where $c_l =
\frac{\delta x}{\delta t}$ is the lattice speed.  Each discrete
velocity $c_i$ is associated with a corresponding discrete
distribution function, $f_i(\vec x,t) \equiv f(\vec x,t;\vec p_i)$ for
the fluid density, and $g_i(\vec x,t) \equiv g(\vec x,t;\vec p_i)$,
with $\vec{p}_i = m \gamma_i \vec{c}_i$, for the fluid
energy-momentum.

The distributions evolve according to the following discrete BGK
Boltzmann equations \cite{BGK}:
\begin{subequations}\label{BGKeq}
  \begin{equation}
    f_i(\vec{x}+\vec{c}_i\delta t,t+\delta t) - f_i(\vec{x},t) =
    -\frac{\delta t}\tau (f_i - f_i^{\rm eq}) \quad ,
  \end{equation}
  \begin{equation}
    g_i(\vec{x}+\vec{c}_i\delta t,t+\delta t) - g_i(\vec{x},t) =
    -\frac{\delta t}\tau (g_i - g_i^{\rm eq}) \quad ,
  \end{equation}
\end{subequations}
where $\tau$ represents the local relaxation time, $f_i^{\rm eq}$ and
$g_i^{\rm eq}$ are the equilibrium distribution functions.

The above equations describe a two-step lattice dynamics.  The
left-hand side encodes the free-streaming of the distributions along
the characteristics defined by the discrete velocities $\vec{c}_i$.
Note that since velocities are constant in space and time, this term
is an {\it exact} lattice transcription of the free-streaming term in
the continuum Boltzmann equation. This leads to major benefits for the
computational performance of the model because, at variance with
hydrodynamic formulations, the information always travels along
straight lines rather than along space-time changing material fluid
lines.

The right-hand side describes particle collisions in the form of a
relaxation towards local equilibria, on a time scale $\tau$.  This is
the lattice analogue of the relativistic Marle model
\cite{MarleModel}.

In order for the relativistic LB equations (Eq.~\eqref{BGKeq}) to
correctly reproduce relativistic hydrodynamics in the continuum limit,
the lattice equilibria have to be designed in compliance with the
basic number-energy-momentum conservation laws.  As shown in
Ref.~\cite{RLB2}, this can be accomplished through an algebraic
moment-matching procedure, leading to the following expressions:

\begin{equation}
\label{EQUIL}
  \begin{aligned}
    f_i^{\rm eq} &= w_i n \gamma \left( 1 + 3\frac{(\vec{c}_i\cdot\vec{u})}{c_l^2}\right) \quad ,\\
    g_0^{eq} &= 3 w_0 P \gamma^2 \left( 4 - \frac{2 +
        c_l^2}{\gamma^2 c_l^2} -2\frac{|\vec{u}|^2}{c_l^2} \right) \quad ,\\
    g_{i>0}^{eq} &= 3 w_i P \gamma^2 \left( \frac{1}{\gamma^2
        c_l^2} + 4\frac{(\vec{c}_i\cdot\vec{u})}{c_l^2} + 6
      \frac{(\vec{c}_i\cdot\vec{u})^2}{c_l^4} - 2
      \frac{|\vec{u}|^2}{c_l^2} \right) \quad ,
  \end{aligned}
\end{equation}

The above probability distribution functions recover the macroscopic values,
with the ultra-relativistic state equation $\epsilon = 3P$, provided
the following identifications are made:

\begin{align}\label{macros}
  n &= \frac{1}{\gamma} \sum_i f_i \quad ,\\
  P &= -\frac{1}{3} \sum_{i} g_i + \frac{1}{3}\sqrt{-3 \left(\sum_{i}
      g_i\vec{c}_i\right)^2 +
    4\left(\sum_{i} g_i\right)^2} \quad , \\
  \vec{u} &= \frac{1}{3 \sum_i g_i + 3 P} \sum_i g_i \vec{c}_i \quad . \\
\end{align}

In the above, the discrete weights take the value $w_0 = 1/3$, for
$|\vec{c}_0|=0$, $w_i = 1/18$ for $|\vec{c}_i|=c_l$, and $w_i = 1/36$
for $|\vec{c}_i|=\sqrt{2} c_l$.  Based on these expressions, it is
readily checked that $\sum_i w_i = 1$ and $c_s^2 = \sum_i w_i c_{ia}^2
= \frac{c_l^2}{3}$, where $a=x,y,z$.  The latter defines the sound
speed, $c_s$, and shows that, by stipulating $c_l =c$, i.e. the
lattice speed equal to the light speed, the RLB supports the
relativistic ideal equation of state $c_s^2=\frac{c^2}{3}$.

It is worth emphasizing that, unlike local equilibria in continuum
momentum space, lattice equilibria are not unconditionally positive
for any value of the fluid velocity.  This is because continuum
equilibria, both non-relativistic (Maxwell-Boltzmann) and relativistic
(Maxwell-Juettner), are irrational functions of both the microscopic
and hydrodynamic velocity (four-momentum). This is no accident, but
rather the result of the local equilibria following from an entropy
minimization principle, with the entropy additivity imposing a
non-rational (exponential) functional dependence on the collision
invariants \cite{LIBOFF}. Hence, an exact transcription of such
equilibria would require an infinite series in powers of $u/c_s$.
However, since the generic $n$-th term of such an expansion involves
$n$-th order tensors of the form $\sum_i \vec{c}_i \vec{c}_i \dots
\vec{c}_i$, it is clear that much larger symmetry groups, i.e. much
more discrete velocities, would be needed at each step of the
expansion.  This would rapidly lead to an unmanageable complexity.  It
is quite fortunate that hydrodynamics can be reproduced by ensuring
the correct symmetry of just fourth-order tensors, so that, relatively
simple lattices like the D3Q19, can accomplish the task.  Failing such
fortunate circumstance, no LB would ever exist.

To model shock waves in viscous quark-gluon matter, this scheme has to
recover a special viscosity-entropy density ratio $\frac\eta s$, where
the entropy density $s$ is approximated by $s=4n-n\ln\lambda$, with
$\lambda = \frac{n}{n^{eq}}$ and $n^{eq} = \frac{d_G T^3}{\pi^2}$
($d_G = 16$ is the degeneration for gluons and $T$ is the temperature)
\cite{qcshockwave}.

The dynamic viscosity can be computed from the relaxation time
according to the following expression:

\begin{equation}
\label{VISCO}
\eta = \frac{4}{3} c_l^2 \gamma P(\tau - \frac{\delta t}{2})
\end{equation}

The term $c_l^2 (\tau - \frac{\delta t}{2})$ is standard from
conventional Lattice Boltzmann theory. However, in contrast to the
classical Lattice Boltzmann method, this term is pre-factored by $\gamma
P/c^2$ ($c=1$ in our units) rather than by fluid density $\rho$.

For the purpose of this work, $\tau$ is computed with the
initial $P$ and $\gamma$, so as to obtain the desired $\frac\eta s$
ratio. Note that the initial $P$ and $\gamma$, in general for the
Riemann problem, are functions of the coordinates, so they can lead
to a spatially dependent relaxation time $\tau$. To avoid this, we
have used their spatial averages.

\subsection{Numerical validation}

\begin{figure}[]
  \centering
  \subfigure[Pressure]{\label{subfig:p21}\includegraphics[width=0.42\textwidth]{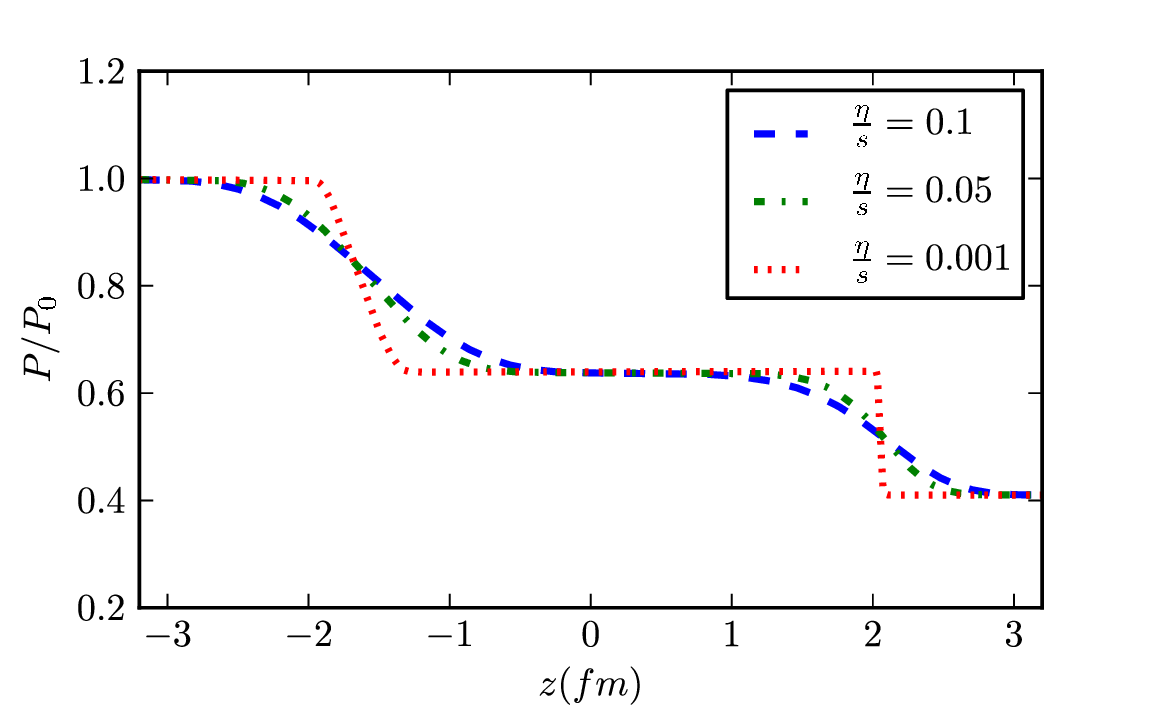}} \\
  \subfigure[Velocity]{\label{subfig:u21}\includegraphics[width=0.42\textwidth]{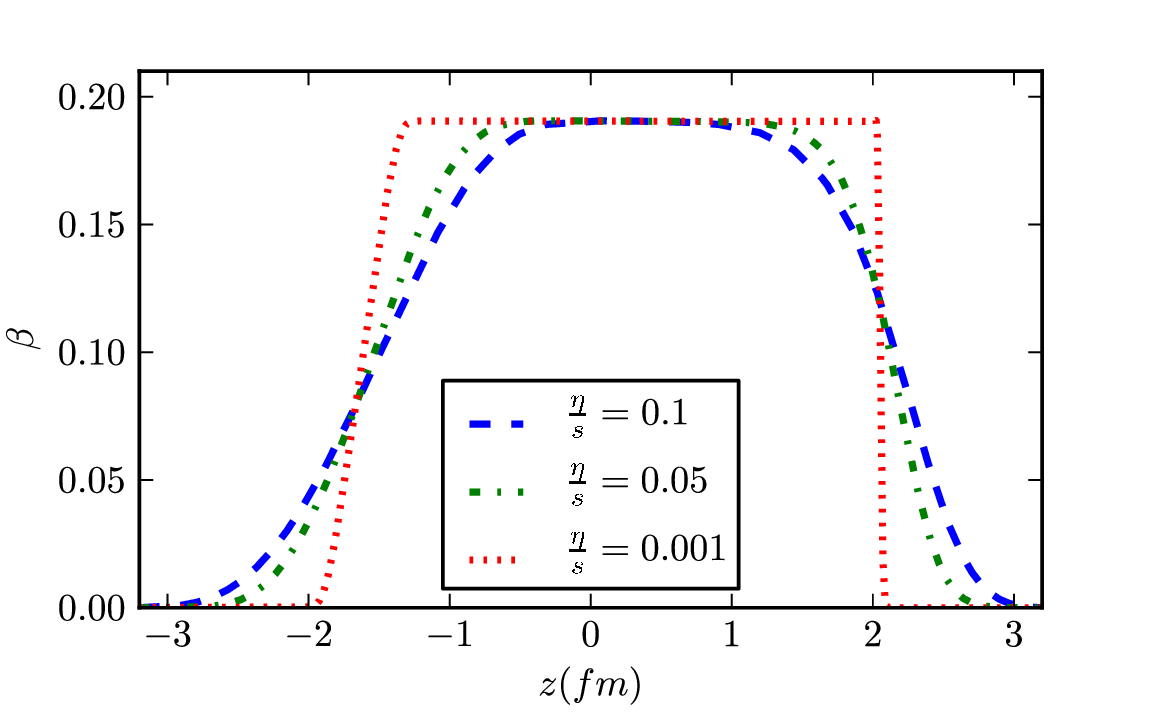}}
  \caption{Pressure and velocity profiles for different $\frac\eta s$
    ratios, at $t = 3.2 \frac{fm}{c}$. The profiles show excellent
    agreement with the results in Ref.\cite{RLB2}.}
  \label{fig:21}
\end{figure}
In order to test our numerical scheme, we carry out some simulations
of shock waves in quark-gluon plasma\cite{qcshockwave} in one
dimension, and compare the results with the existing literature,
Ref.~\cite{RLB2}. For this simulation, a lattice with
$1\times1\times800$ cells, open boundaries along the mainstream
$z$-direction and periodic boundaries along the cross-flow $x$-and
$y$-directions are used.  As a result, we set $\delta x = 0.008 fm$
and $\delta t = 0.008 \frac{fm}{c}$. The initial conditions for the
pressure are $P(z<0)=P_0=5.43 \frac{GeV}{fm^3}$ and $P(z\ge0)=P_1 =
2.22 \frac{GeV}{fm^3}$. This corresponds, in numerical units, to
$2.495\times10^{-7}$ and $1.023\times10^{-7}$, respectively.  The
initial temperature is constant over the whole domain having the value
$350MeV$ (in numerical units $0.0314$), corresponding to $\zeta \sim
3$.

The initial particle density is computed through the relation $n=\frac
PT$.  The pressure and velocity profiles at $3.2\frac{fm}{c}$ for
different $\frac\eta s$ ratios between $0.001$ and $0.1$, are shown in
Fig.~\ref{fig:21}, from which excellent agreement with the results in
Ref.\cite{RLB2} is readily appreciated. Each simulation took around
one second on a Intel core i5 of 2.3GHz.

\section{Improving the equilibrium distribution}
\label{sec:improvement}
\begin{figure}[]
  \centering
  \includegraphics[width=0.42\textwidth]{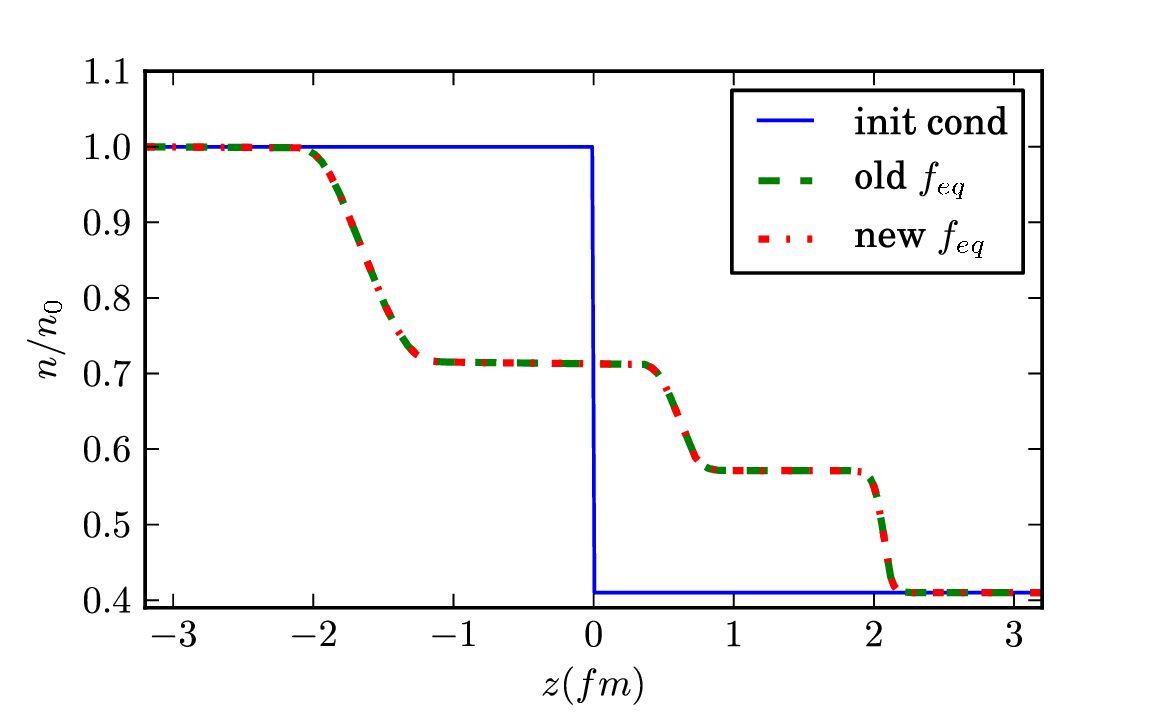}
  \caption{Comparison of the particle density using both equilibrium
    functions, Eqs.~\eqref{EQUIL} and \eqref{newEQUIL}, at $t =
    3.2\frac{fm}c$, with the initial conditions given in
    section~\ref{sec:initcond1}, $\frac\eta s = 0.005$, and $c_l = 1$.
    As one can appreciate, the density profiles are basically the
    same.}
  \label{fig:n2x}
\end{figure}
\begin{figure}
  \centering
  \includegraphics[width=0.42\textwidth]{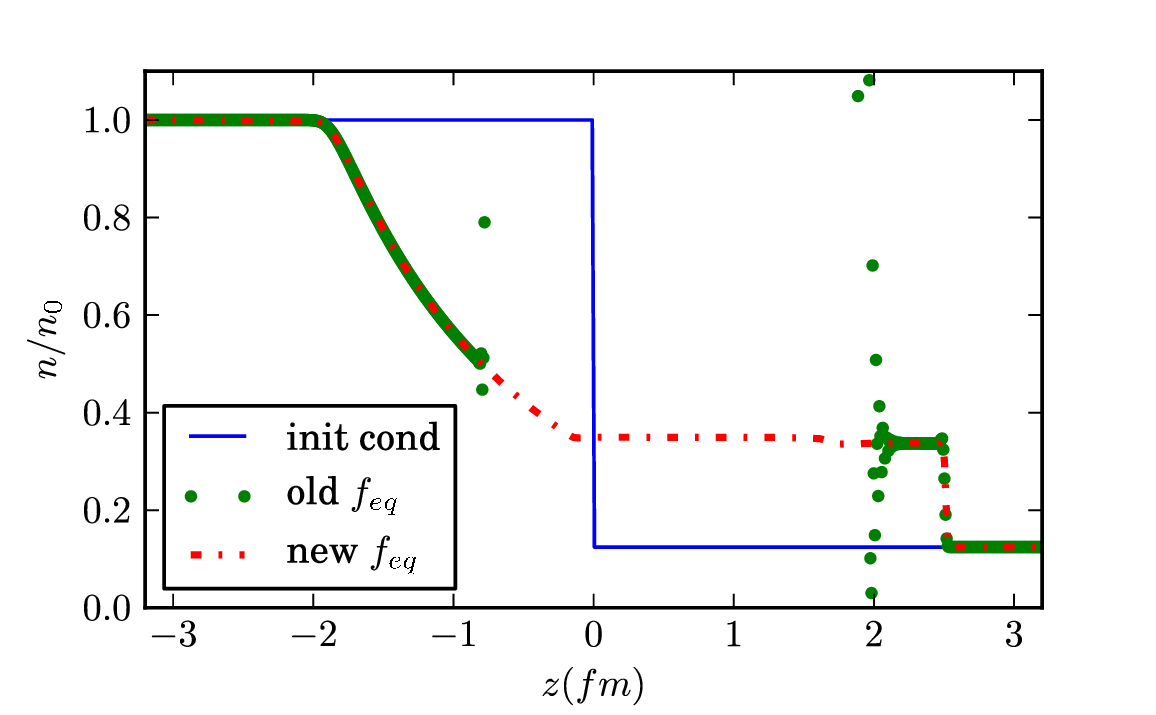}
  \caption{Comparison of the particle density using both equilibrium
    functions, at $t = 3.2\frac{fm}c$, with the initial conditions
    given in section~\ref{sec:initcond2}, $\frac\eta s = 0.001$, and
    $c_l = 1$. The result obtained with the old (linear) equilibrium
    function breaks down in the region $z \in [-1,2]$, whereas for the
    case of the new (quadratic) equilibrium function, the model
    reproduces the correct profile in the whole domain.}
  \label{fig:n6x}
\end{figure}

Inspection of the particle density equilibrium distribution function,
$f_i^{\rm eq}$$=$$w_i n \gamma \left( 1 + 3 \frac{(\vec{c}_i \cdot
    \vec{u})}{c^2_l} \right)$, shows that this becomes negative for
counter-streaming populations when $(\vec{c}_i\cdot\vec{u}) <-
\frac{c_l^2}{3}$.  This is non-physical and leads to either unreliable
results (see Fig.~\ref{fig:n6x}) or numerical instability.

In previous works \cite{RLB1,RLB2}, this problem was circumvented by
increasing the lattice speed $c_l$ and reducing the time step
accordingly, so that the light-cone condition $c_l \delta t = \delta
x$, remained fulfilled.  In the low-viscosity regime this was found to
stabilize the numerics without affecting the physics to any
appreciable extent.  However, as shown in the sequel, in at higher
viscosities, this is no longer the case.

A detailed analysis of the effect of the lattice speed on the physical
results is presented in section~\ref{sec:infcl}.  However, prior to
this, we first introduce a new particle density equilibrium function,
which allows larger fluid velocities at a given value $c_l=1$, without
violating the positivity condition, $f_i^{\rm eq} >0$.

This new equilibrium distribution reads as follows:
\begin{equation}\label{newEQUIL}
  f_i^{\rm eq} = \omega_i n \gamma \left(1 + 3\frac{(\vec{c}_i \cdot
      \vec{u})}{c_l^2} + \frac92\frac{(\vec{c}_i \cdot
      \vec{u})^2}{c_l^4} - \frac32\frac{|\vec{u}|^2}{c_l^2}\right) \quad .
\end{equation}

This expression includes a new quadratic term in the macroscopic fluid
velocity. This new term is standard in classical lattice Boltzmann
theory, and results form second order expansion in the fluid velocity
of a locally shifted Maxwellian distribution $\sim
e^{-\frac{m(v-u)^2}{2kT}}$.

To check that the new equilibrium function is valid and reproduces the
same results as the old one, we carry out several simulations in the
weakly relativistic regime, which each of them required around one
second on an Intel core i5 of 2.3GHz. For illustration purposes, we show
just one example, where we only consider the particle distribution
function, because the improved equilibrium function has no direct
influence neither on the pressure nor on the velocities, see
Eq.~\eqref{macros}.

From Fig.~\ref{fig:n2x}, which shows that density profile at
$\eta/s=5\; 10^{-3}$, we observe that both equilibrium distribution
functions give basically the same result. However, from
Fig.~\ref{fig:n6x}, which refers to $\eta/s=10^{-3}$, we can see that
the computation of the particle density using the old equilibrium
distribution function, in the moderately relativistic regime, is
breaking down in the region $z \in [-1,2]fm$, and leads to very large
fluctuations, with both positive and negative values. For
visualization purposes, just the physically meaningful region is
plotted. Note that outside of the region $z \in [-1,2]fm$ both
equilibrium functions reproduce the same result.  However, in the
course of the evolution, those fluctuations are found to propagate
over the entire simulation domain.

\section{Influence of the lattice speed}
\label{sec:infcl}
Having illustrated the stabilization effects of the new equilibrium
distribution, all following numerical experiments are performed with
this distribution. Next, we study the influence of the lattice speed,
$c_l$. For this purpose, we simulate various shock waves in
quark-gluon plasma with different viscosity-entropy density ratios and
different lattice speeds in the weak ($1< \gamma \ll 2$) and mildly
($\gamma \sim 2$) relativistic regimes.  As noted before, by
increasing the lattice speed, we need to decrease the time step, in
such a way that $\vec{x} + \vec{c}_i dt$ corresponds to a grid point
in the lattice. Basically, in numerical units, $c_l = 10$ implies
$dt=0.1$, so that the product is always $1$ or $\sqrt{2}$, depending
of the velocity vector $\vec{c}_i$.  Clearly, smaller time-steps imply
a correspondingly larger number of time steps, hence more simulation
time.

\begin{figure}[]
  \includegraphics[width=0.42\textwidth]{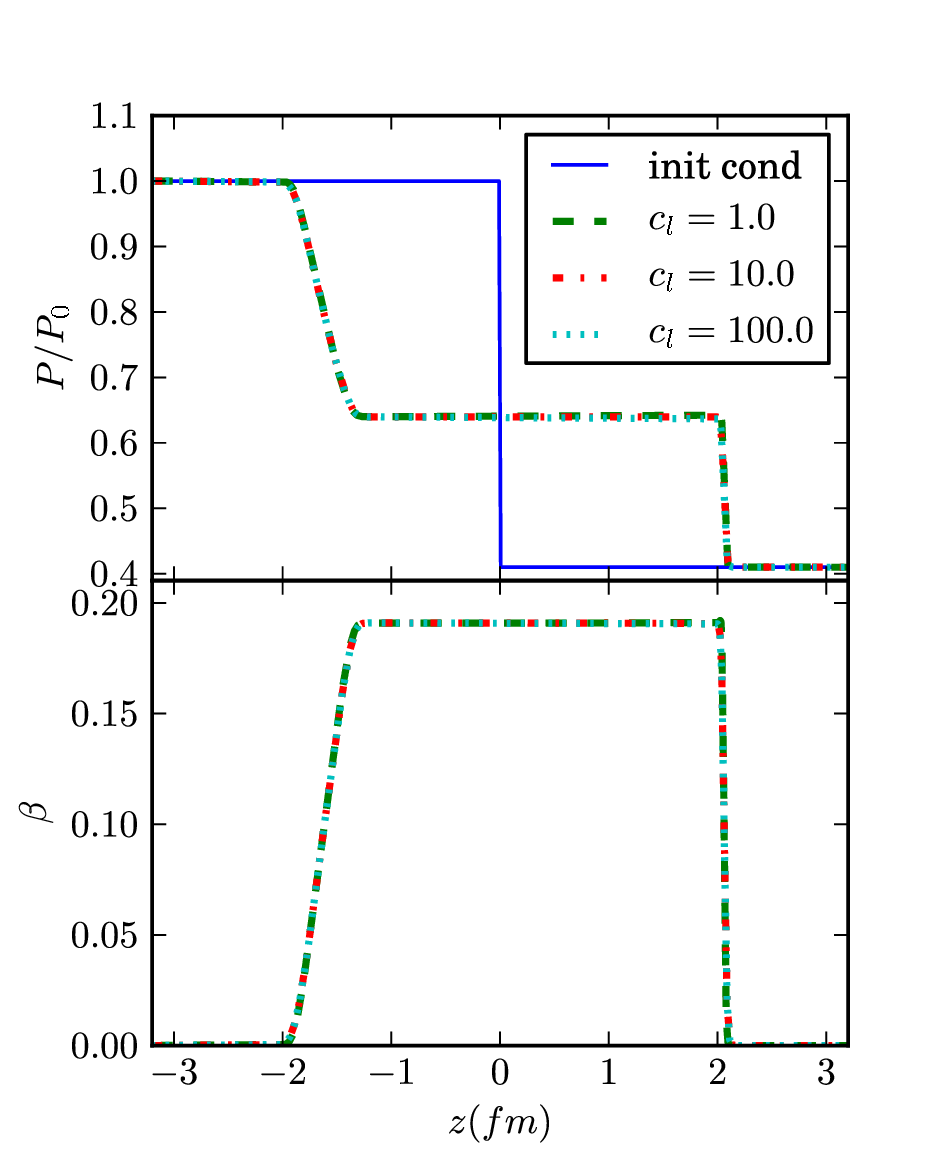}
  \caption{Weakly relativistic case, $\frac{\eta}{s}=0.001$.  Pressure
    and velocity profiles for different lattice speeds at $t =
    3.2\frac{fm}c$. In this case, the choice of the lattice speed does
    not affect the result of the simulation.}  \label{fig:b2e1}
\end{figure}
\begin{figure}[]
  \includegraphics[width=0.42\textwidth]{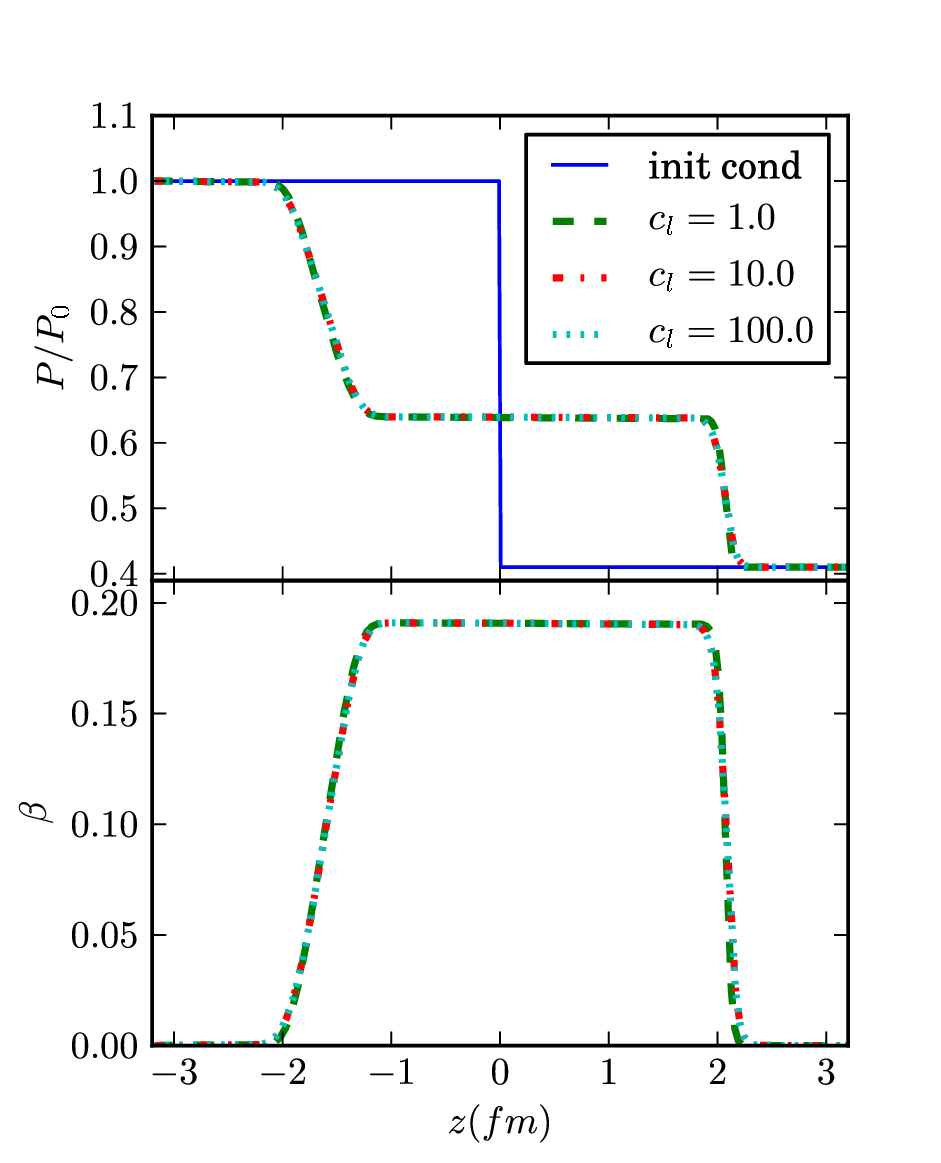}
  \caption{Weakly relativistic case, $\frac{\eta}{s} = 0.005$.
    Pressure and velocity profiles for different lattice speeds and ,
    at $t = 3.2\frac{fm}c$. Here, some differences are visible.}
  \label{fig:b2e5}
\end{figure}
\begin{figure}[]
  \includegraphics[width=0.42\textwidth]{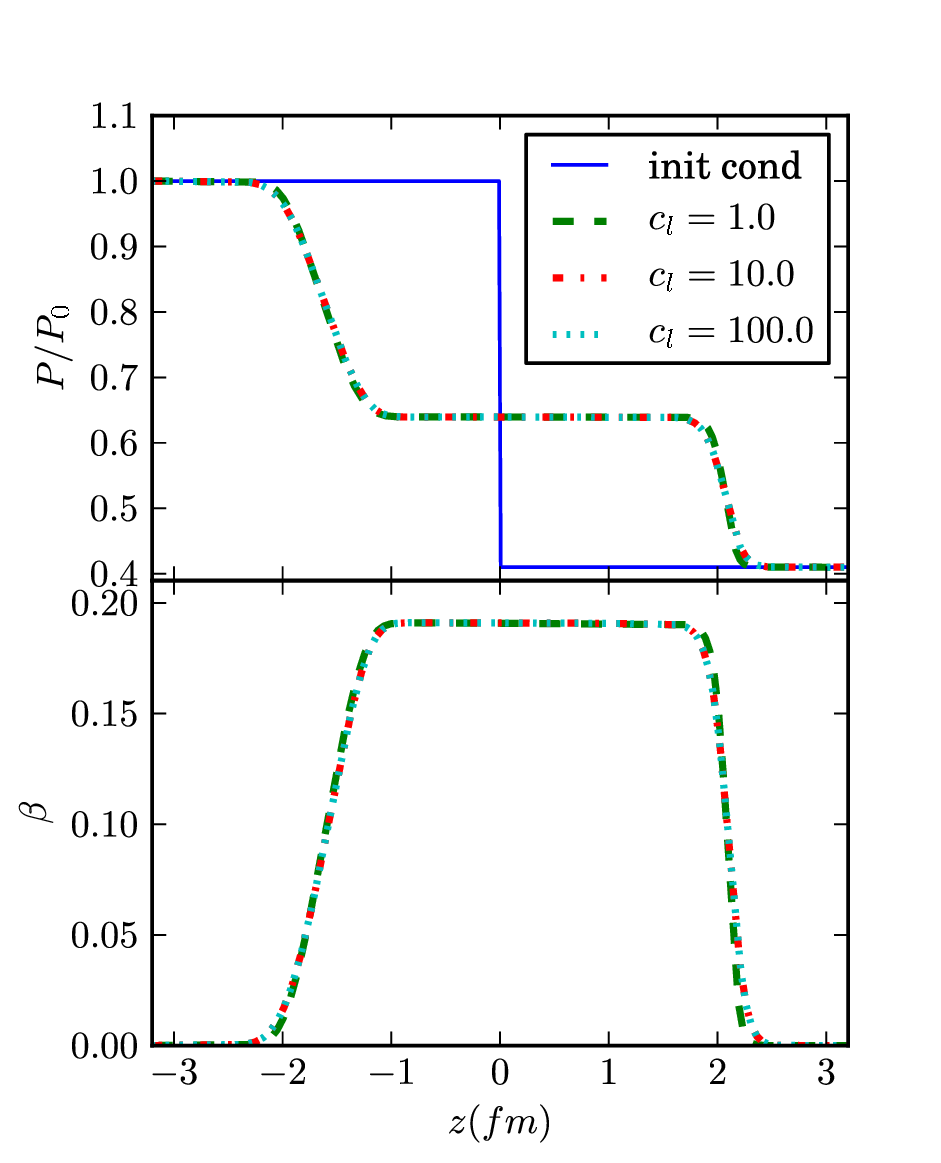} 
  \caption{Weakly relativistic , $\frac{\eta}{s} = 0.01$.  Pressure
    and velocity profiles for different lattice speeds at $t =
    3.2\frac{fm}c$. Here we observe a small effect of the lattice
    speed.}
  \label{fig:b2e10}
\end{figure}

\subsection{Weakly relativistic regime, $\gamma \ll 2$}
\label{sec:initcond1}

For the weakly relativistic simulations, we use a lattice with
$1\times1\times 800$ cells, open boundaries in $z$-direction and
periodic boundaries in $x$-and $y$-direction.  We set $\delta x =
0.008 fm$ and $\delta t = \frac{0.008}{c_l} \frac{fm}{c}$.  The
initial conditions for the pressure are $P(z<0)=P_0 = 5.43
\frac{GeV}{fm^3}$ and $P(z \ge 0) = P_1 = 2.22 \frac{GeV}{fm^3}$. This
corresponds in numerical units to $2.495\times10^{-7}$ and
$1.023\times10^{-7}$, respectively. The initial temperature is
constant over the whole domain, namely $400MeV$(in numerical units
$0.036$), corresponding to $\zeta \sim 2.5$.  The initial particle
density is calculated with the equation of state, $n=\frac pT$.

A snapshot of the pressure and velocity profiles at $t =
3.2\frac{fm}{c}$, for different lattice speeds($c_l=1$, $c_l=10$, and
$c_l = 100$) and different viscosity-entropy ratios (between $0.001$
and $0.01$), is shown in Figs. \ref{fig:b2e1}, \ref{fig:b2e5} and
\ref{fig:b2e10}.  From these figures, we can observe that , as
expected based on previous experience, there is no significant effect
of the lattice speed on the results. However, there is a significant
difference in the computational performance, the simulations took
around $1$, $9$, and $90$ seconds for $c_l = 1.0$, $c_l = 10.0$ and
$c_l = 100.0$, respectively.

\subsection{Moderately relativistic regime, $\gamma \sim 2$}
\label{sec:initcond2}

For the moderately relativistic regime, we perform the simulations on
a lattice with $1\times1\times1600$ cells, in order to cover a larger
computational domain. The same boundaries are used as before.  We set
$\delta x = 0.008fm$ and $\delta t = \frac{0.008}{c_l} \frac{fm}{c}$.
The initial conditions for the pressure are $P_0=5.43\frac{GeV}{fm^3}$
and $P_1=0.339\frac{GeV}{fm^3}$. In numerical units, they correspond
to $2.495\times10^{-7}$ and $1.557\times10^{-8}$, respectively.  The
initial temperature is now different for $z \ge 0$ namely $T_1 = 200
MeV$ (in numerical units $0.018$), corresponding to $\zeta \sim 5$.

\begin{figure}[]
  \includegraphics[width=0.42\textwidth]{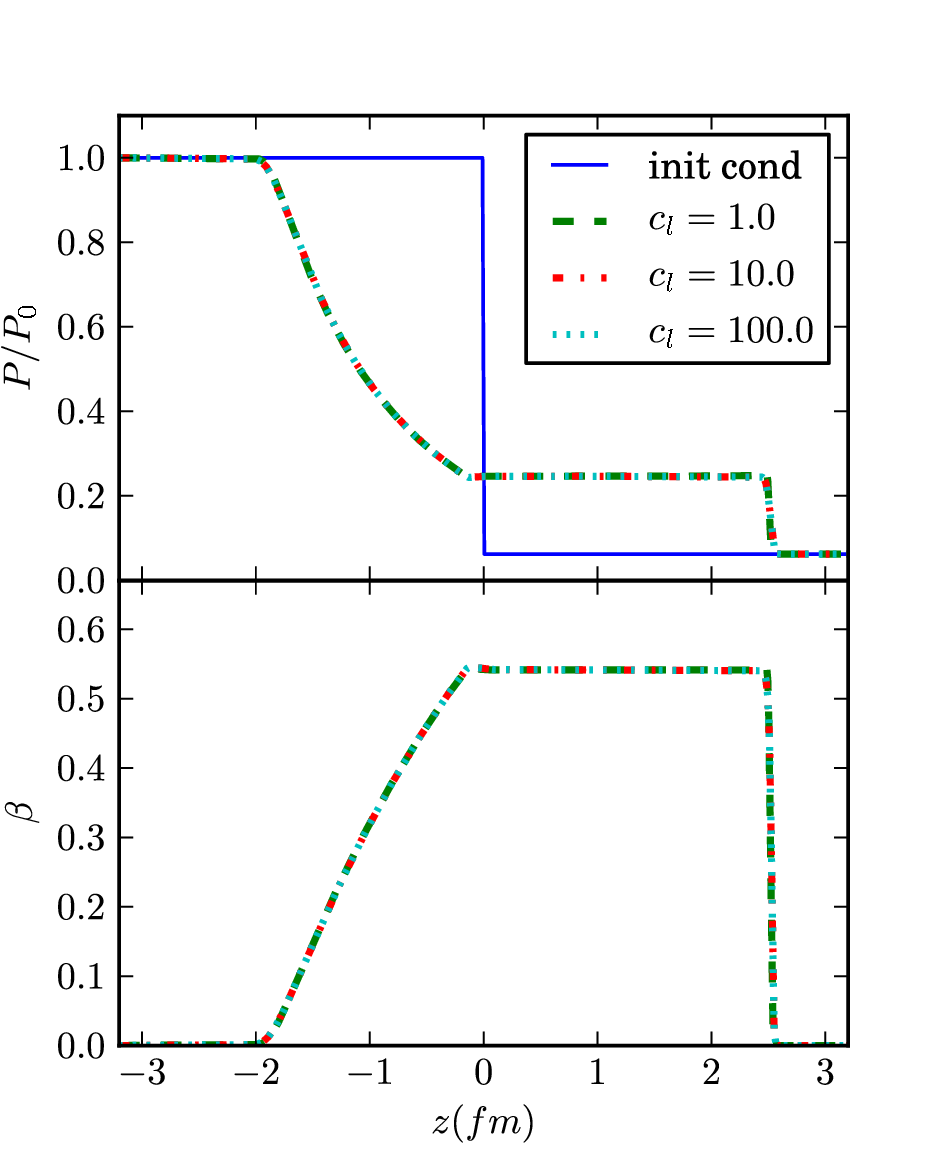}
  \caption{Moderately relativistic case, $\frac{\eta}{s} = 0.001$.
    Pressure and velocity profiles for different lattice speeds at $t
    = 3.2\frac{fm}c$. Very small differences appear only at the
    highest $\beta$.}\label{fig:b6e1}
\end{figure}

\begin{figure}[]
  \includegraphics[width=0.42\textwidth]{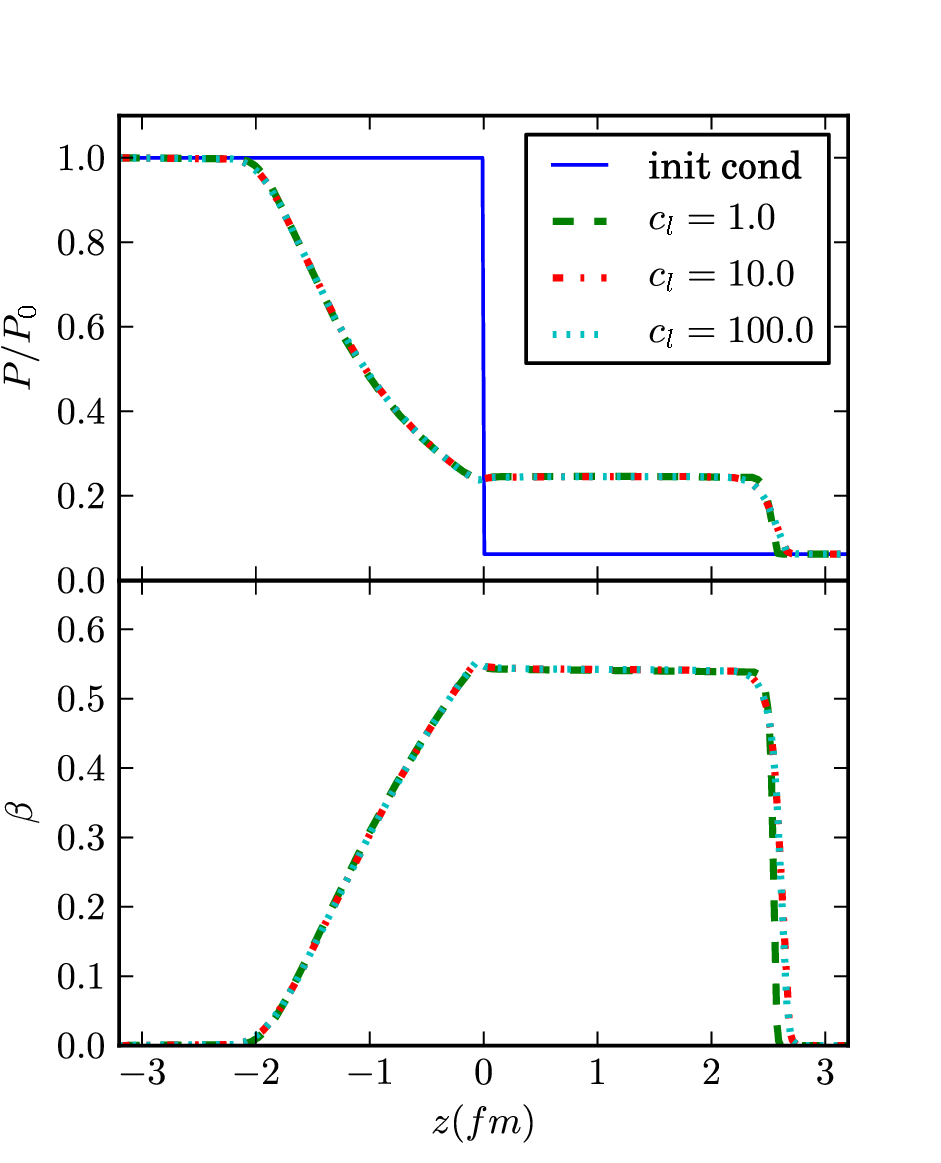}
  \caption{Moderately relativistic case, $\frac{\eta}{s} = 0.005$.
    Pressure and velocity profiles for different lattice speeds at $t
    = 3.2\frac{fm}c$.  Here we see some differences for different
    lattice speeds.}\label{fig:b6e5}
\end{figure}

For $z < 0$ the initial temperature is the same as in the previous
simulation, $T_0=400MeV$ (in numerical units $0.036$), and the initial
particle density is computed in the same way as before.  The velocity
and pressure profiles at $t = 3.2\frac{fm}c$, for different lattice
speeds($c_l = 1$, $c_l=10$, and $c_l = 100$) and different
viscosity-entropy density ratios ($0.001$ and $0.05$), are shown in
Figs. ~\ref{fig:b6e1} and~\ref{fig:b6e5}.  Again, no significant
influence of the lattice speed is observed.

\begin{figure}[]
  \includegraphics[width=0.42\textwidth]{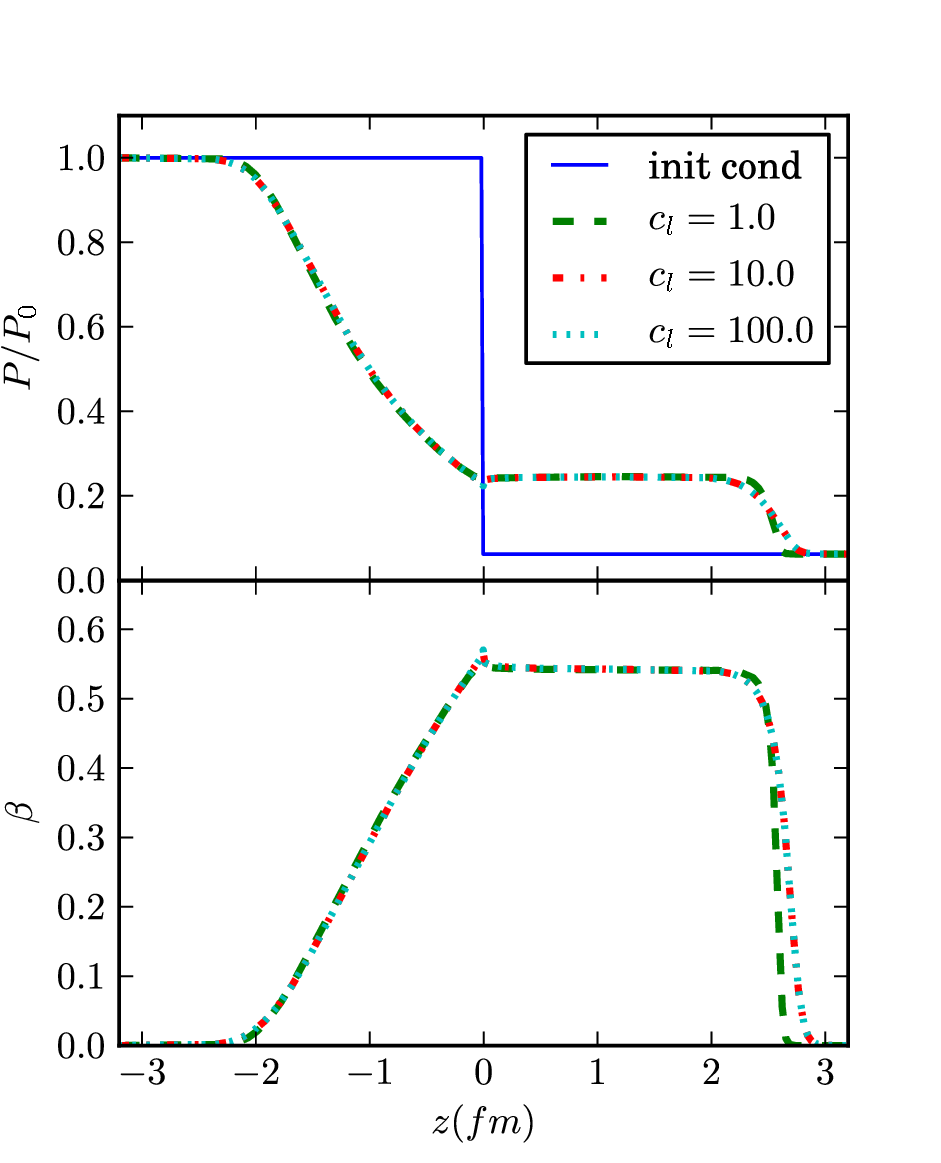}
  \caption{Moderately relativistic case, $\frac{\eta}{s} = 0.01$.
    Pressure and velocity profiles for different lattice speeds at $t
    = 3.2\frac{fm}c$.  Here we observe, that especially for the
    highest $\beta$, the results differ
    substantially.}\label{fig:b6e10}
\end{figure}

For $\frac\eta s=0.01$, we perform the simulation using a lattice with
$1\times1\times800$ cells, so that $\delta x=0.016fm$ and $\delta
t=\frac{0.016}{cl} \frac{fm}c$.  In this case, the numerical units of
the pressure are $P_0 = 1.996\times10^{-6}$ and $P_1 =
1.2456\times10^{-7}$, respectively. The numerical values of the
temperatures and particle density are set up in the same way as
before. The results of this simulation are shown in Fig.
~\ref{fig:b6e10}.  Here, we see that the lattice speed drastically
affects the results. The implemented simulation in this section
spanned around $1$, $4$, and $44$ seconds for the cases $c_l = 1.0$,
$c_l = 10.0$, and $c_l = 100.0$, respectively.

Summarizing, we conclude that the physical results in the moderately
relativistic regime are affected by the increasing value of the
lattice speed beyond $c_l = 1.0$. Also, we see that this effect is
more pronounced with increasing $\frac\eta s$. Furthermore, we observe
that the results of the simulation with high lattice speeds, for a
given ratio $\frac\eta s$, are similar to the ones obtained with $c_l
= 1$ for a higher viscosity-entropy density ratio.
To understand this effect, a deeper theoretical investigation of the
basic RLB scheme is required.  According to the Chapman-Enskog
expansion, the relation for $\eta$ results from a Taylor
expansion of the discrete lattice Boltzmann equation to second order
in space and time, combined with a first order perturbation in the
Knudsen number $Kn \sim c_s \tau/\delta x = \frac{1}{3} \tau/\delta
t$, and quadratic truncation of the local equilibria in the Mach
number $Ma=u/c_s = \sqrt 3 \beta$.  Since, as we have observed before,
raising $c_l$ implies lowering $\delta t$, it is clear that
simulations with $c_l>1$ at a given $\tau$, imply larger values of the
Knudsen number, thus pushing RLB potentially outside of the domain of
validity of the Chapman-Enskog asymptotics.  Similar problems are well
known to occur in the simulation of complex non-relativistic fluids
with sharp interfaces \cite{sharp1}.


\section{Comparison with BAMPS and vSHASTA for high values of $\frac\eta s$}

\begin{figure}[]
  \centering
  \subfigure[Pressure]{\label{subfig:compp}\includegraphics[width=0.42\textwidth]{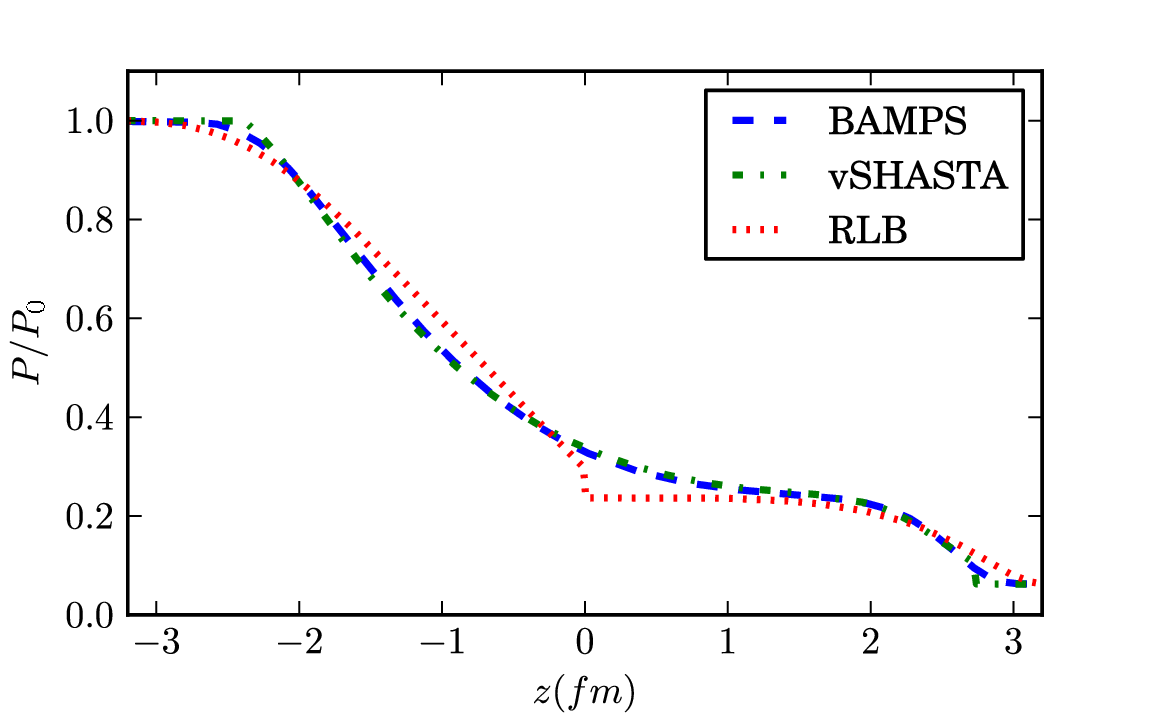}} \\
  \subfigure[Velocity]{\label{subfig:compu}\includegraphics[width=0.42\textwidth]{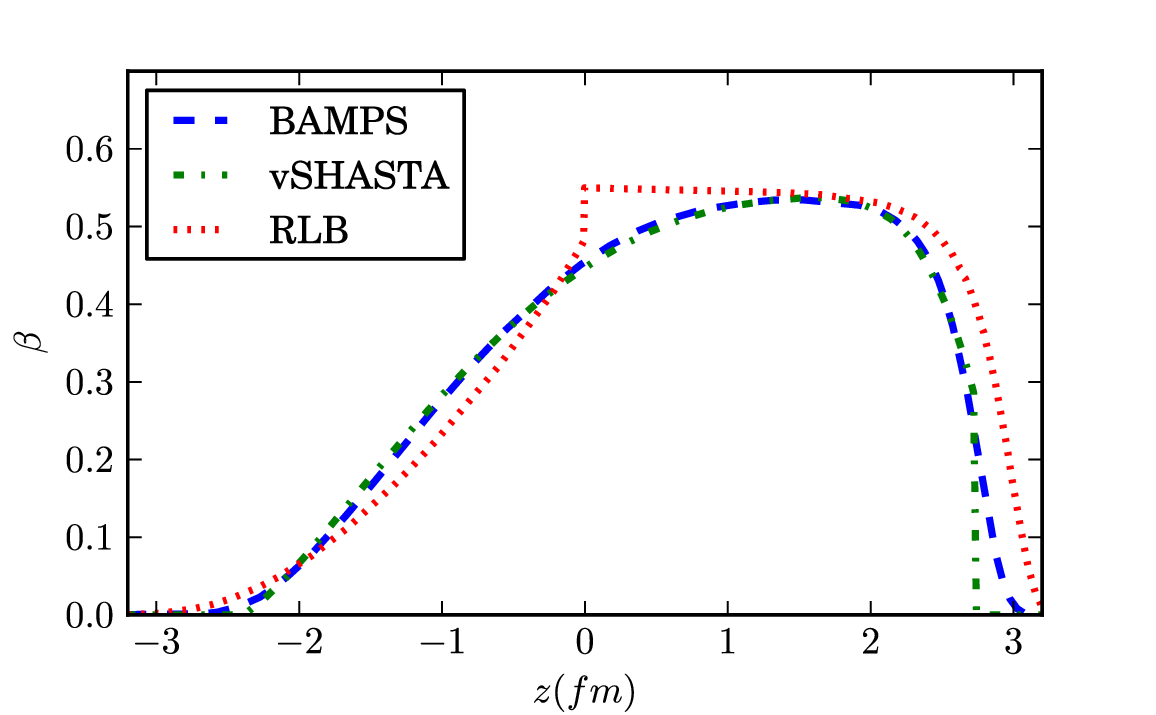}}
  \caption{Pressure and velocity profiles for the different models
    with $\frac{\eta}{s} = 0.1$.  BAMPS and vSHASTA show only tiny
    departures from each other, but differ significantly from RLB.  }
  \label{fig:comp1}
\end{figure}

\begin{figure}[]
  \centering
  \subfigure[Pressure]{\label{subfig:compp5}\includegraphics[width=0.42\textwidth]{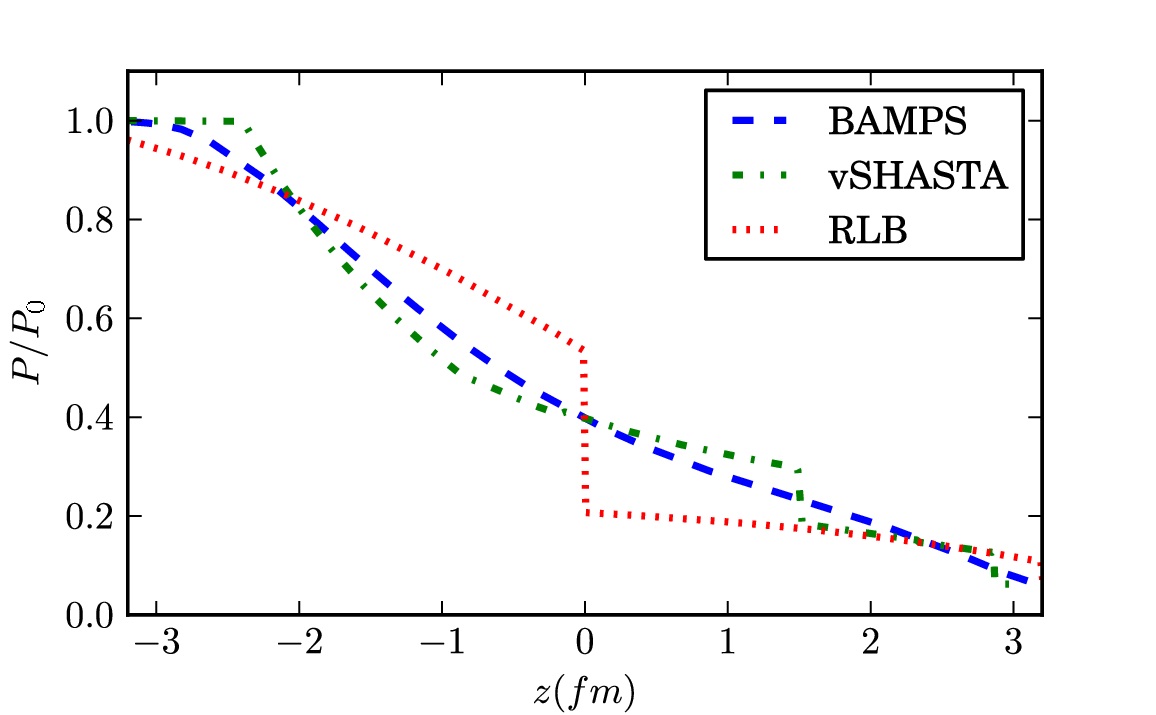}} \\
  \subfigure[Velocity]{\label{subfig:compu5}\includegraphics[width=0.42\textwidth]{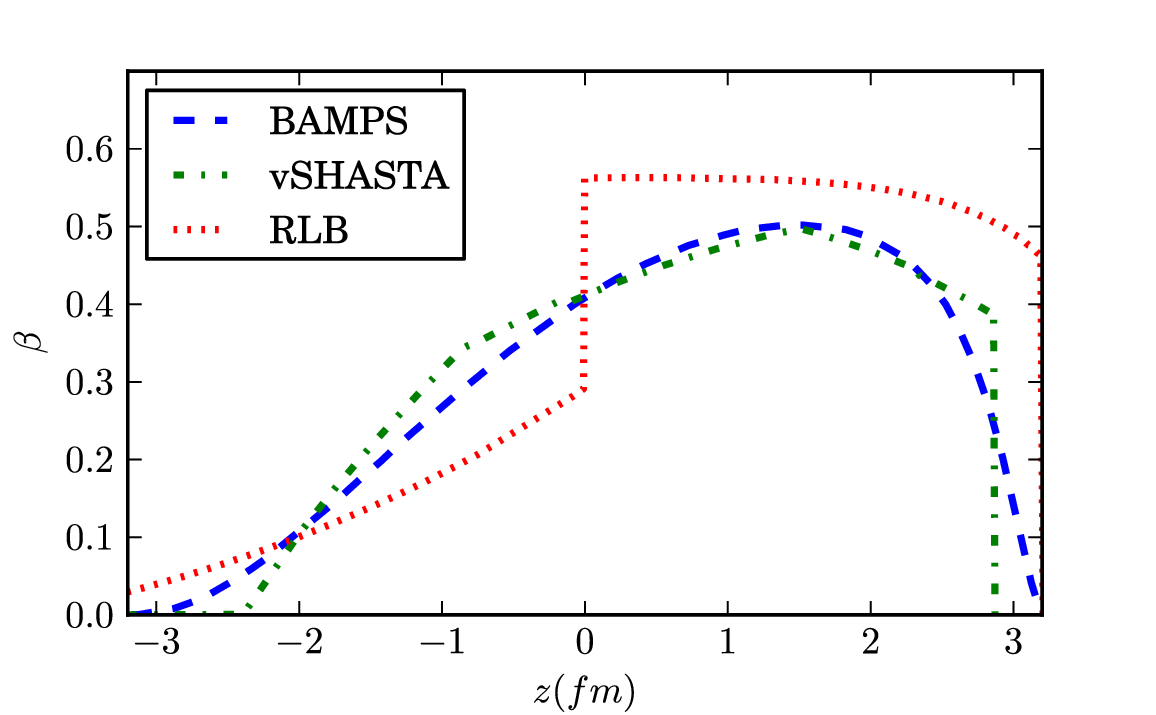}}
  \caption{Pressure and velocity profiles for the different models
    with $\frac{\eta}{s} = 0.5$. vSHASTA has stability problems and
    discontinuities, while the RLB diverges from the BAMPS solution
    even more.}
  \label{fig:comp5}
\end{figure}

In this section, we perform simulations to compare the RLB model with
BAMPS and vSHASTA \cite{data} for the case of high viscosities and
high speeds. The initial conditions are the same as in
section~\ref{sec:initcond2}. We use $c_l = 1.0$ and a lattice with
$1\times1\times1600$ cells, so that the numerical units stay the same.
Each simulation took around one second on an Intel core i5 of 2.3GHz.
The results of the simulation at $t = 3.2\frac{fm}c$ for two different
viscosities are shown in Figs.~\ref{fig:comp1} and~\ref{fig:comp5}.
Due to the fact that BAMPS is the solution of the Boltzmann equation
with the complete collision term, while the RLB model is a
near-equilibrium approximation, and vSHASTA is a second order scheme
for modeling relativistic fluid dynamics, it is very interesting to
compare the three techniques with each other.

For $\frac\eta s = 0.1$ the results from BAMPS and vSHASTA are very
close, and the RLB presents some small deviations.  On the other hand,
for $\frac\eta s = 0.5$, BAMPS still reproduces stable results,
whereas vSHASTA suffers some stability problems and discontinuities,
and the deviations of the RLB become more pronounced. To understand
these deviations in high viscosity regime, we start first by
revisiting the meaning of the viscosity-entropy density ratio within
the RLB model.  The viscosity-entropy density ratio can be written as
\begin{equation}
  \frac\eta s = \frac{\frac43\gamma P \left(\tau-\frac{\delta
        t}2\right) c_l^2}{n\left(4-\ln\left(\frac{n \pi^2}{d_G
          T^3}\right)\right)} \quad ,
\end{equation}
and using the relation $T=\frac P n$,
\begin{equation}
  \frac\eta s = \frac{\gamma c_l^2\left(\tau - \frac{\delta t}2\right)}{3 +
    \frac94\ln\left(\sqrt[3]{\frac{d_G}{n\pi^2}}T\right)} T\quad .
\end{equation}
This expression shows that, for high temperatures, there is a nearly
linear dependence of $\frac\eta s$ on the temperature, which means
that when we simulate a high $\frac\eta s$ ratio, this corresponds, in
terms of the Boltzmann equation, to a high equilibrium temperature.

Note that the results of the RLB simulations present a discontinuity
at $z = 0$ fm in Figs.~\ref{fig:comp1} and \ref{fig:comp5}. This
discontinuity appears as a consequence of the initial condition for
the pressure (solid line in Fig.~\ref{fig:b6e10}), which is set
according to the Riemann problem. Therefore, we can conclude that the
RLB, in the case of high viscosity-entropy density ratios, is not able
to solve correctly the Riemann problem and maintains the initial
discontinuity during the whole simulation.

\begin{figure}[h!]
  \centering
  \includegraphics[width=0.41\textwidth]{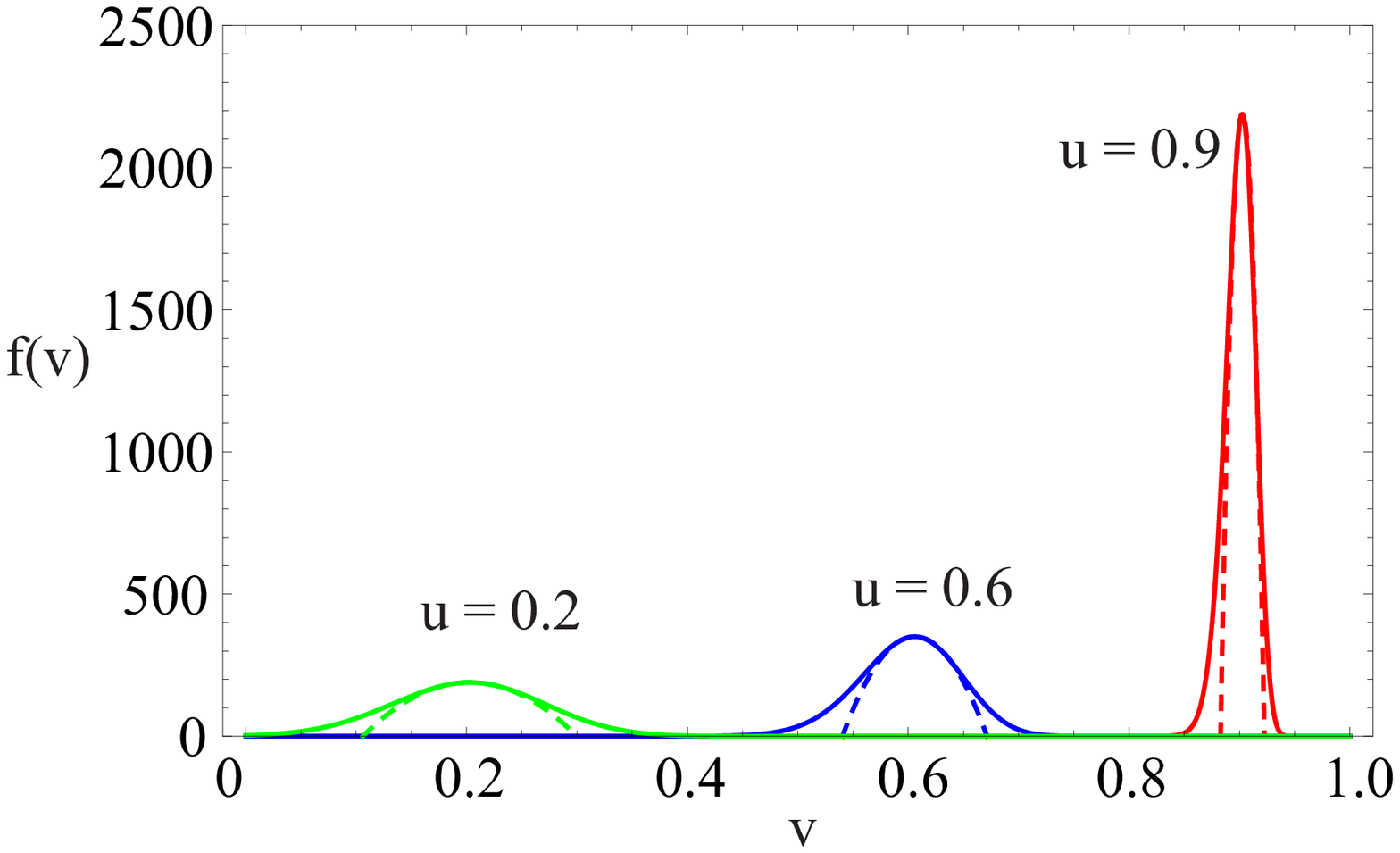}
  \caption{Maxwell-J\"uttner distribution for different velocities at
    very low temperatures, $m c^2 /kT = 200$, the dashed lines denote
    the parabolic approximation to the respective MJ distribution,
    denoted by solid lines. Only the positive velocities are shown
    since the distribution is symmetric.}
  \label{fig:mjlt}
\end{figure}

The RLB model approximates the probability equilibrium distribution
functions by a quadratic expression, i..e a parabolic approximation.
At sufficiently low temperature, this approximation is accurate, but
as temperature is increased, hence higher values $\frac\eta s$,
accuracy is rapidly lost.  To appreciate this effect, it is
instructive to graphically inspect the Maxwell-J\"uttner distribution
(MJ)\cite{RBE1} $f_{MJ}(\vec x,\vec p,t) = \frac1Z
\exp\left(-\frac{U^\alpha p_\alpha}T\right)$, which depends on the
macroscopic four-velocity $U^\alpha = \gamma(\vec{u})\left( 1 ,
  \vec{u} \right)$, and the microscopic four-momentum $p_\alpha =
\gamma(\vec v) m\left(1 ,-\vec v \right)$, where $m$ is the rest mass
and $\vec v$ the microscopic velocity and $Z$ a normalization factor
that depends on the temperature and macroscopic velocity.

\begin{figure}[h!]
  \centering
  \includegraphics[width=0.41\textwidth]{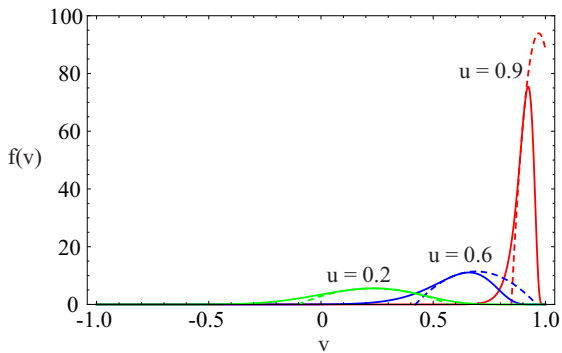}
  \caption{Maxwell-J\"uttner distribution for different velocities at
    low temperatures, $mc^2/kT = 20$, the dashed lines represent the
    parabolic approximation to the respective MJ distribution, denoted
    by solid lines.}
  \label{fig:mjmt}
\end{figure}

\begin{figure}[h!]
  \centering
  \includegraphics[width=0.41\textwidth]{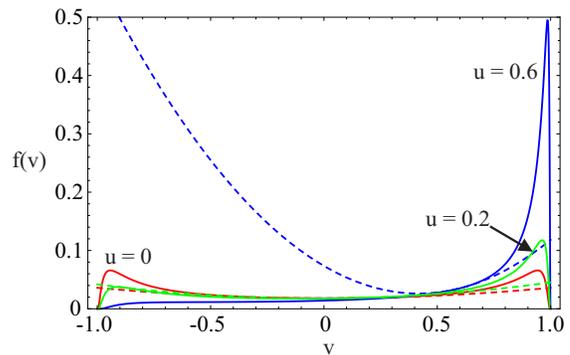}
  \caption{Maxwell-J\"uttner distribution for different velocities at
    high temperatures, $mc^2/kT = 1$, the dashed lines denote the
    parabolic approximation to the corresponding MJ distribution (solid
    lines).}
  \label{fig:mjht}
\end{figure}

From Figs. \ref{fig:mjlt}, \ref{fig:mjmt} and \ref{fig:mjht}, we
observe that the MJ distribution becomes broad and very asymmetric for
high velocities and high temperatures, and therefore a parabolic
approximation is no longer valid. In fact, at $\zeta < 1$, the
parabolic approximation should be replaced by a bimodal parabolic one,
which is certainly feasible, but beyond the current RLB formulation.
This explains why RLB has problems to reproduce results in this
regime. However, for the case of low temperature and relatively high
velocities, the RLB model can still model the proper fluid dynamics
since its equilibrium distribution functions (parabolic approximation)
remain very close to the MJ distribution.  For high temperatures, it
is only possible to study weakly relativistic systems (see
Fig.~\ref{fig:mjht}).

\begin{figure}[h!]
  \centering
  \includegraphics[width=0.32\textwidth]{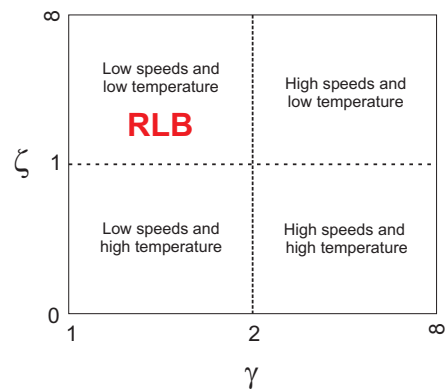}
  \caption{Qualitative sketch of the four different regimes
    characterized by the parameters $\zeta = m c^2 /kT$, and $\gamma =
    1/\sqrt{1-\vec{u}^2}$.  Our RLB model lies in the region of low
    temperatures, $\zeta>1$ and $\gamma < 2$.}
  \label{fig:scheme}
\end{figure} 

As a general conclusion, we expect that the RLB approach can model
moderately-relativistic fluid dynamics ($|\vec{u}| \approx 0.9$) only
in the low temperature regime, $\zeta >1$. For illustration purposes,
we show a qualitative sketch of the regimes characterized by the
parameters $\zeta$ and $\gamma$ in Fig.~\ref{fig:scheme}.

\begin{acknowledgments}
  The authors are grateful to H. Niemi for providing data from vSHASTA
  and to the Center for Scientific Computing (CSC) at Frankfurt
  University for the computing resources. IB is grateful to HGS-Hire.

  This work was supported by the Helmholtz International Center
  for FAIR within the framework of the LOEWE program
  launched by the State of Hesse.
\end{acknowledgments}

\section{Conclusion}

Summarizing, in this paper we have determined the capabilities and
limitations of the RLB model for the simulation of relativistic flows
in the various regimes associated to the flow speed and temperature.
For this purpose, we have performed extensive simulations of shock
waves in quark-gluon plasma at weakly and moderately relativistic
regimes and low and high viscosities. We have introduced a higher
order equilibrium distribution function, which is shown to enhance the
stability of the original RLB model, and permits to compute correctly
the particle density profile in the moderately relativistic regime,
with no need of enhancing the lattice speed, hence no need of reducing
the time-step, as in the original RLB model. The result is that RLB
can compute well-resolved weakly and moderately one-dimensional
relativistic shock wave propagation in less than a minute, CPU time on
an Intel core i5 of 2.3Ghz. To the best of our knowledge, this is
significantly faster than any hydrodynamic code, let alone the full
BAMPS solution.

Furthermore, we have performed a numerical investigation of the effect
of the lattice speed on the physical results, and concluded that the
results can differ, due to higher order terms in the dissipation
tensor which are not included in the Chapman-Enskog analysis. As a
consequence, we observed that the results of the simulation with high
lattice speeds, for a given ratio $\frac\eta s$, are similar to the
ones obtained with $c_l = 1$ for a higher viscosity-entropy density
ratio.

In this study, we also showed that the choice of high
viscosity-entropy density ratios at high speeds, affects the accuracy
of the results. In addition, by direct comparison of the current RLB
model with BAMPS and vSHASTA in a moderately relativistic and highly
viscous regime, we have shown that the parabolic approximation of the
Maxwell-Juettner distribution poses significant restrictions to the
viability of RLB for strongly relativistic, high temperature, fluids.

Finally, based on the study of the approximation of the MJ
distribution with parabolic equilibria, we are led to predict, that
the current RLB model would properly work at relatively low
temperatures, $\zeta > 1$, and high velocities ($\beta \sim 0.9$).
Fortunately, such regimes are by no means devoid of interesting
physical applications.  Extension to more general relativistic flows
requires further developments, such as the introduction of higher
order lattices, with higher order equilibria, possibly equipped with
relativistic H-theorems (entropic LB methods).  This offers a very
interesting object of future research in the field.

\bibliography{main}

\end{document}